\documentclass[preprint,12pt]{elsarticle}
\usepackage{amssymb}

\usepackage[T1]{fontenc}

\journal{Bio-Algorithms and Med-Systems}

\begin{document}

\begin{frontmatter}

\title{Deep learning model for ECG reconstruction reveals the information content of ECG leads.}

\author[1]{Tomasz Gradowski}
\ead{tomasz.gradowski@pw.edu.pl}
\author[1]{Teodor Buchner}

\affiliation[1]{organization={Warsaw University of Technology, Faculty of Physics},
            addressline={Koszykowa 75}, 
            city={Warsaw},
            postcode={00662}, 
            country={Poland}}
\begin{abstract}
This study introduces a deep learning model based on the U-net architecture to reconstruct missing leads in electrocardiograms (ECGs). The model was trained to reconstruct 12-lead ECG data from reduced lead configurations using publicly available datasets. The results highlight the ability of the model to quantify the information content of each ECG lead and its inter-lead correlations. This has significant implications for optimizing lead selection in diagnostic scenarios, particularly in settings where complete 12-lead ECGs are impractical. In addition, the study provides insights into the physiological underpinnings of ECG signals and their propagation. The findings pave the way for advances in telemedicine, portable ECG devices, and personalized cardiac diagnostics by reducing redundancy and improving signal interpretation.
\end{abstract}

\begin{keyword}
Electrocardiogram (ECG) \sep Lead Reconstruction \sep Convolutional neural network (CNN) \sep  U-Net \sep Deep Learning \sep Machine Learning
\end{keyword}

\end{frontmatter}


\section{Introduction}
\label{sec:introduction}

The electrocardiogram (ECG) is a widely used diagnostic tool in cardiology. It is a non-invasive method to record the electrical activity of the heart. The standard ECG consists of 12 leads that are recorded from different positions on the body. Each lead provides a different view of the heart's electrical activity, and the combination of all leads gives a comprehensive picture of the heart’s function. However, in some cases, not all leads are available due to technical limitations or the patient’s condition.
Furthermore, reducing the number of leads reduces patient burden and device complexity. It therefore improves the availability of ECG equipment and the data quality \cite{Yoo2023}, which defines a clear design target \cite{Xue2024}. In such cases, it is desirable to be able to reconstruct the missing leads from the available leads. However, an information theory aspect is also crucial apart from this technical motivation. Specifically, concepts such as redundancy (the presence of overlapping information across leads), mutual information (a measure of shared information content between signals), and reconstruction fidelity (the ability to regenerate missing signals from partial observations) underlie our study.

Various methods have been explored to evaluate the information content of ECG electrodes. Linear correlation is the most commonly used method \cite{Pipberger1961-fi, Holderith2018-ft, Jain2023-rg}, which also enables detecting electrode reversal [6] and allows evaluating the possibilities of regeneration of 12 leads from a custom reduced electrode set, such as proprietary EASI \cite{Holderith2018-ft} or corrected orthogonal leads \cite{Pipberger1961-fi}. A correlation matrix may be obtained, either by classical correlation \cite{Jain2023-rg} or with the use of dedicated Deep Neural Networks (DNN) \cite{Zhang2022-le}, which allows one to study the extent of overlap between the data provided by various electrodes. Electrodes of minimal overlap are considered the most important for ventricular arrhythmia detection \cite{Zhang2022-le}. The vast capabilities offered by modern neural networks also allow reconstruction of missing data between electrodes or between time segments \cite{Lence2025}. 

However, it has long been known that correlation, or lack thereof, is of paramount importance to vectorcardiography, which was historically based on the initial assumption that the limb and the precordial leads represent projection onto various directions in the frontal and transverse planes, respectively.  
Vectorcardiography relies on the assumption that the limb leads and the precordial leads provide approximately orthogonal projections of the electrical activity of the heart in the frontal and transverse planes, respectively. In this context, the degree of correlation between leads is critical: Low correlation indicates greater spatial independence, which improves the fidelity and interpretability of vector representations of the cardiac electrical field. This principle is based on the classical vectorcardiography theory \cite{Frank1956}.

The idea of validation of the current reference point for precordial leads - the Wilson central terminal \cite{Gargiulo, Gargiulo2016}, determination of an optimum reference point \cite{Moeinzadeh} verification of the added value delivered either by vectorcardiography alone \cite{Man2015} or by its specific representations \cite{Boonstra} easily follows.

Modern DNNs present a unique opportunity to study not only simple linear correlations but also the true added value of the information content between various electrode setups. It is essential to understand that each of the correlations observed in the ECG stems from physical relations present in the significant, spatially extended source, i.e., the heart.
The first point of the analysis is the analysis of limb leads: only two of them are statistically independent, as I + III = II. Does it mean that, in the spatial direction of lead II, we observe combined activity from two other spatial directions: lead I and lead III? The same relations are obtained for Goldberger augmented leads, a linear combination of limb leads. Therefore, lead III and augmented leads are excluded from further analysis, as their relationship will be strong and directly follow their definition.

The second point is the relationship between the limb and precordial leads. Here, the situation is more subtle, as we observe a mirror effect of composed limb leads, which is reflected in precordial leads, to which it serves as a reference. The analysis of mirror patterns has a long tradition \cite{Schmitt, Brody}, but it is also studied in a contemporary context \cite{Vaidya, Wang}.

The third point concerns the interrelations between precordial leads. Due to the small distance between electrodes, the difference in signal is not so significant, and an apparent gradual change of QRS polarity is visible \cite{Meek}, which is a clear correlation between neighboring electrodes. Electrical shunting by excessive amounts of electrode gel is a common mistake, which completely abolishes the observed changes \cite{Riera}.

Anyway, this short resume shows that various relations concerning ECG leads' information content are known and that the utilization of DNN allows the extension of the statistical models far beyond the Pearson correlation coefficient. Apart from motivation from basic science, many practical aspects were already mentioned: an ability to reduce the electrode set or vice versa, to utilize the over-completeness of the ECG to cover missing spatiotemporal segments, and last but not least, to determine the electrodes that are crucial from the point of view of arrhythmia analysis. It is important to note that due to its bipolar nature, the ECG does represent only a spatial gradient, which is only a difference between much stronger unipolar signals. Therefore, care must be taken to maximize the information carried by this signal, which represents the asymmetry of cardiogenic electromagnetic wave \cite{Buchner}.

This study presents a deep learning model to reconstruct missing ECG leads. The model is based on the U-net architecture \cite{unet}, which is effective for image segmentation tasks. We train the model on a large dataset of ECG recordings and evaluate its performance on a separate validation dataset. We show that the model can reconstruct missing leads and that the reconstruction quality depends on the source leads.
A key advantage of the model is that it can estimate the amount of information carried by each lead and the correlations between leads. This information can help understand the heart’s underlying physiology and improve the accuracy of ECG interpretation.
The remainder of the paper is organized as follows. In Section 2, we describe the methods used in the study. In Section 3, we present the results of the study. In Section 4, we discuss the results and their implications. Finally, in Section 5, we present our conclusions and suggest directions for future work.

\section{Methods}
\subsection{Model}

\begin{figure}[h]
    \centering
    \includegraphics[width=\textwidth]{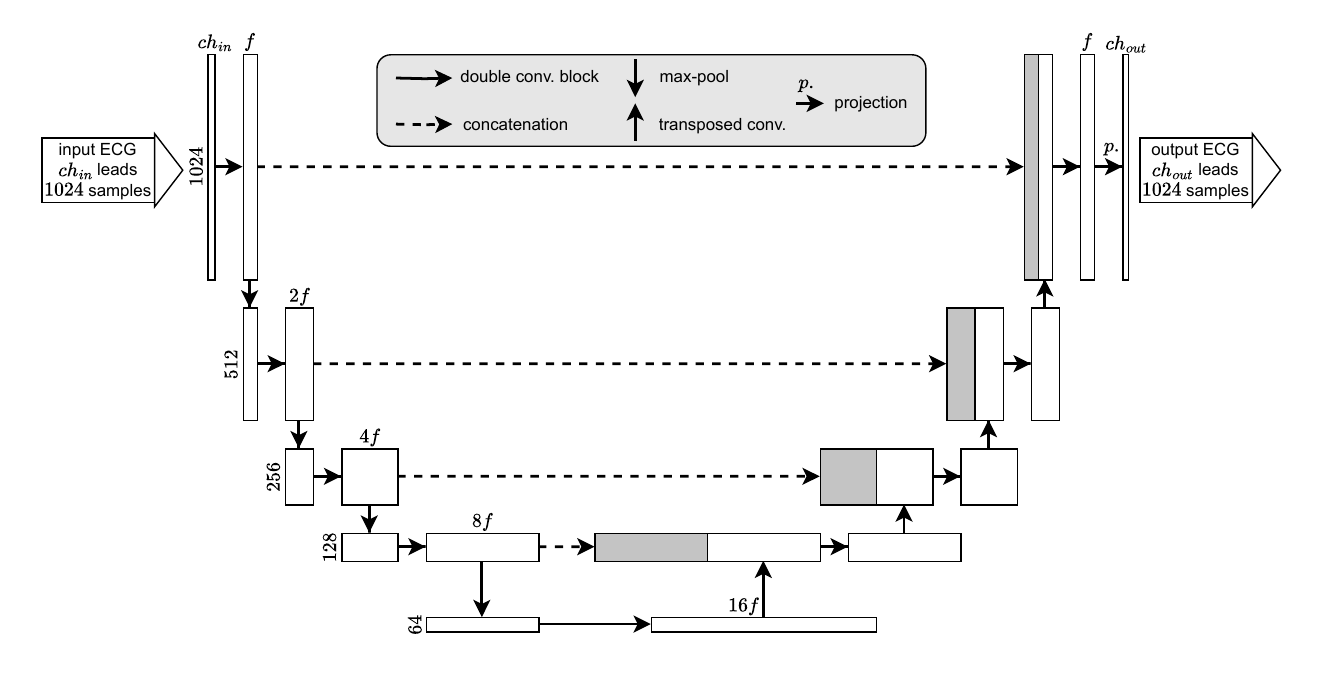}
    \caption{U-net architecture adopted for ECG reconstruction. Rectangular blocks represent tensors with dimensions shown at the top (number of channels) and on the left (number of samples). In our model, the initial number of channels $f=64$ was used. $ch_{in}$ and $ch_{out}$ represent the number of input and output leads, respectively.
    }\label{fig:unet}
\end{figure}

The model used in this study is based on the U-net architecture \cite{unet}, which consists of an encoder-decoder network with skip connections between the corresponding layers of the encoder and decoder. The encoder part of the network consists of a series of convolutional and max-pooling layers that down-sample the input data. The decoder part of the network consists of a series of convolutional and upsampling layers, which upsample the data to the original resolution. The skip connections between the encoder and decoder layers allow the network to capture the input data's local and global features.

We implemented the model as close as possible to the original U-net architecture, with some modifications to adapt it to the ECG reconstruction task (Fig.\ref{fig:unet}). In particular, we use one-dimensional convolution instead of two-dimensional. The model consists of an encoder with four convolutional blocks ({\it conv-blocks}), one conv-block in the bottleneck, and a decoder with four conv-blocks. Each convolutional block consists of two layers with a kernel size of 3, a stride of 1, and padding of 1, followed by a batch normalization layer and a ReLU activation function. The dropout layers, present in the original U-net architecture, were removed from our model for simplicity, since their removal did not significantly affect the model's performance.

The first convolutional layer of each block changes the feature maps' depth. The downsampling in the encoder is done by max-pooling layers with a kernel size of 2 and a stride of 2. Transposed convolutional layers do the upsampling in the decoder with a kernel size of 2 and a stride of 2. The final layer of the network is a convolutional layer with a kernel size of 1 for the projection of the feature maps to the number of leads to be reconstructed. There is no activation function in the final layer.

In selecting the model architecture, our primary goal was not to achieve the absolute best possible reconstruction accuracy, but to enable a meaningful comparison between different input lead configurations. To achieve this, the model had to be sufficiently robust to reconstruct complex ECG signals, but not so overpowered that it would minimize differences between various input sets. A model that is either too weak (high reconstruction errors across all configurations) or too strong (nearly perfect reconstructions regardless of input selection) would obscure the evaluation of lead information content. The U-net architecture provides a balanced compromise: it is simple, interpretable, and powerful enough to achieve reasonable reconstructions, yet sensitive enough to reveal variations in reconstruction quality depending on the available leads. This design choice is analogous to selecting an appropriately challenging exam to distinguish students of varying skill levels—too easy or too difficult a test would fail to differentiate performance meaningfully.

The model takes a set of ECG leads as input and outputs a set of reconstructed leads. The input leads are encoded as a 2D array, where the rows correspond to the time samples, and the columns correspond to the leads. The output leads are also encoded as a 2D array, with the same number of rows as the input leads and a variable number of columns depending on the number of leads to be reconstructed. The length of the input and output leads is set to 1024 samples (8192 ms).

The model is trained using a dataset of ECG recordings. The model is trained to reconstruct missing leads. The loss function used for training is the mean squared error (MSE) between the reconstructed leads and the ground truth, as it is a widely used metric in autoencoder-based models \cite{autoencoders}. The model is trained for 100 epochs using the AdamW optimizer with a learning rate of $3 \cdot 10^{-4}$ and a batch size of 32.

To perform the study, we performed 25 training sessions with different sets of input leads:
3 models with single limb lead (I, II, III) as input,
1 model with two limb leads (I, II) as input,
6 models with two limb leads (I, II) and one chest (precordial) lead (V1, V2, V3, V4, V5, V6) as input,
and 15 models with two limb leads (I, II) and all possible combinations of 2 chest leads as input.

In all cases, the model was trained to reconstruct all precordial leads that were not present in the set of input leads. To accommodate a different number of input and output leads, the number of filters was changed in the first and the last convolutional layer.

\subsection{Data}

The data used in this study comes from two separate datasets, one for training and validation and one for testing. The training and validation sets were based on the PTB-XL dataset \cite{ptbxl1, ptbxl2}, a large publicly available electrocardiography dataset. The dataset contains 21799 12-lead ECG recordings from 18869 patients. The recordings were sampled at 500 Hz and lasted 10 seconds. For this study, we downsampled the recordings to 125 Hz and truncated them to 1024 samples (8192 ms). The dataset was split into a training set and a validation set, with $90\%$ of the recordings used for training and $10\%$ used for validation. No data augmentation techniques were applied during training. The PTB-XL dataset is already highly diverse, containing ECG recordings annotated with a wide range of diagnostic classes, including normal ECG, myocardial infarction, ST/T changes, conduction disturbances, and hypertrophy. This inherent clinical variability provides sufficient coverage of physiological and pathological ECG morphologies for the training task.

The testing dataset was based on the PTB dataset \cite{ptb}, which is a publicly available dataset of 549 12-lead ECG recordings from 290 patients. The recordings were sampled at 1000 Hz, varying from 30 to 120 seconds. For this study, we downsampled the recordings to 125 Hz and split them into 1024-sample (8192 ms) segments, getting 7061 segments in total. PTB-XL and PTB datasets are available from the PhysioNet platform \cite{physionet}.

\section{Results}

\begin{figure}[h]
    \centering
    \includegraphics[width=\textwidth]{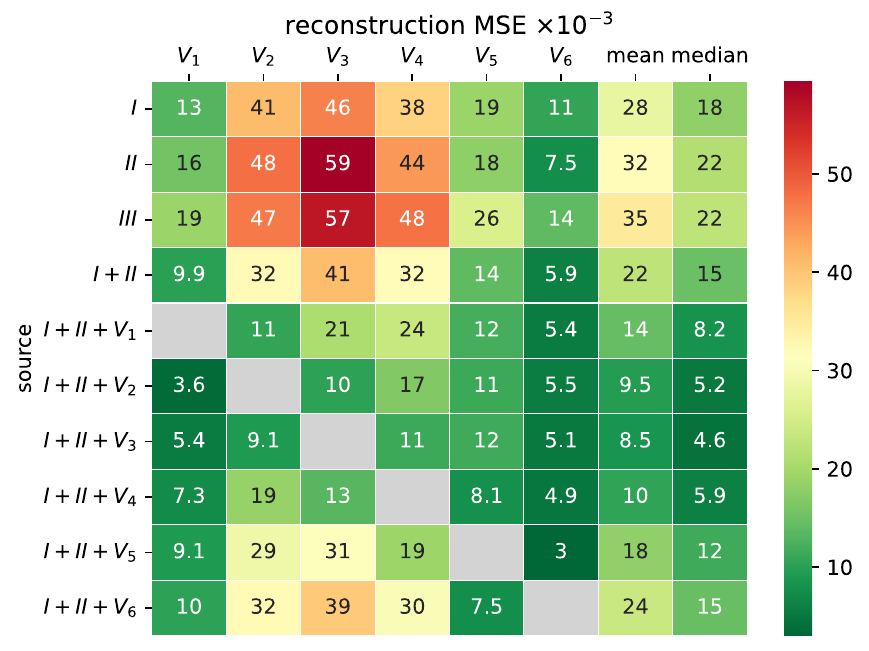}
    \caption{Performance of the models trained with single limb leads as input (I, II, III), two limb leads as input (I, II), and two limb leads and one precordial lead as input (I, II, Vx). The table shows the mean MSE between the reconstructed leads and the ground truth for each model. Mean and median MSE values for the testing dataset are shown in the rightmost columns.}\label{fig:single}
\end{figure}

In this section, we present the results of the study. We show the model's performance on the testing dataset. Our model is, in fact, a set of models, each trained with a different set of input leads. We evaluated the performance of each model by calculating the mean squared error (MSE) between the reconstructed leads and the ground truth. We also show the reconstructed leads for selected examples.

The results are presented in two groups. The first group consists of models trained with one limb lead as input (I, II, III), two limb leads as input (I, II), and two limb leads and one precordial lead as input (I, II, Vx). The second group consists of models trained with two limb leads and two precordial leads as input (I, II, Vx, Vy). In each group, we show the results for selected examples.

Fig. \ref{fig:single} shows the results of the models in the first group. The first three rows show the results of the models trained with one limb lead as input. The fourth row shows the results of the model trained with two limb leads as input. We excluded the model trained with all three limb leads as input, as the leads are linearly dependent ($I + III = II$), and the third lead does not carry any additional information. All reported values in Figure \ref{fig:single} represent the average performance across 7061 ECG segments from the PTB testing dataset. The model predicted the missing leads for each segment, and the reconstruction error was computed individually before averaging.

In addition to the mean reconstruction error, we also report the median MSE for each input configuration. This is important because the distribution of MSE values is not symmetric or Gaussian; a small number of extreme outliers, often due to rare pathological signals or sharp artifacts, significantly inflate the standard deviation. Therefore, standard deviation alone is not a reliable indicator of uncertainty. The mean reflects the average-case performance, while the median reflects the typical-case performance, providing a more robust summary of model behavior under non-normal error distributions.

As shown in Fig. \ref{fig:single}, leads V1 and V6 exhibited the lowest reconstruction errors across all single-limb input models, indicating a stronger correlation with limb lead signals. In contrast, V3 and V4 had consistently higher reconstruction errors, suggesting these leads capture more unique spatial information. Notably, when V3 was included among the input leads, reconstruction performance across all other precordial leads improved significantly, confirming its central informational role.

\begin{figure}[h]
    \centering
    \includegraphics[width=\textwidth]{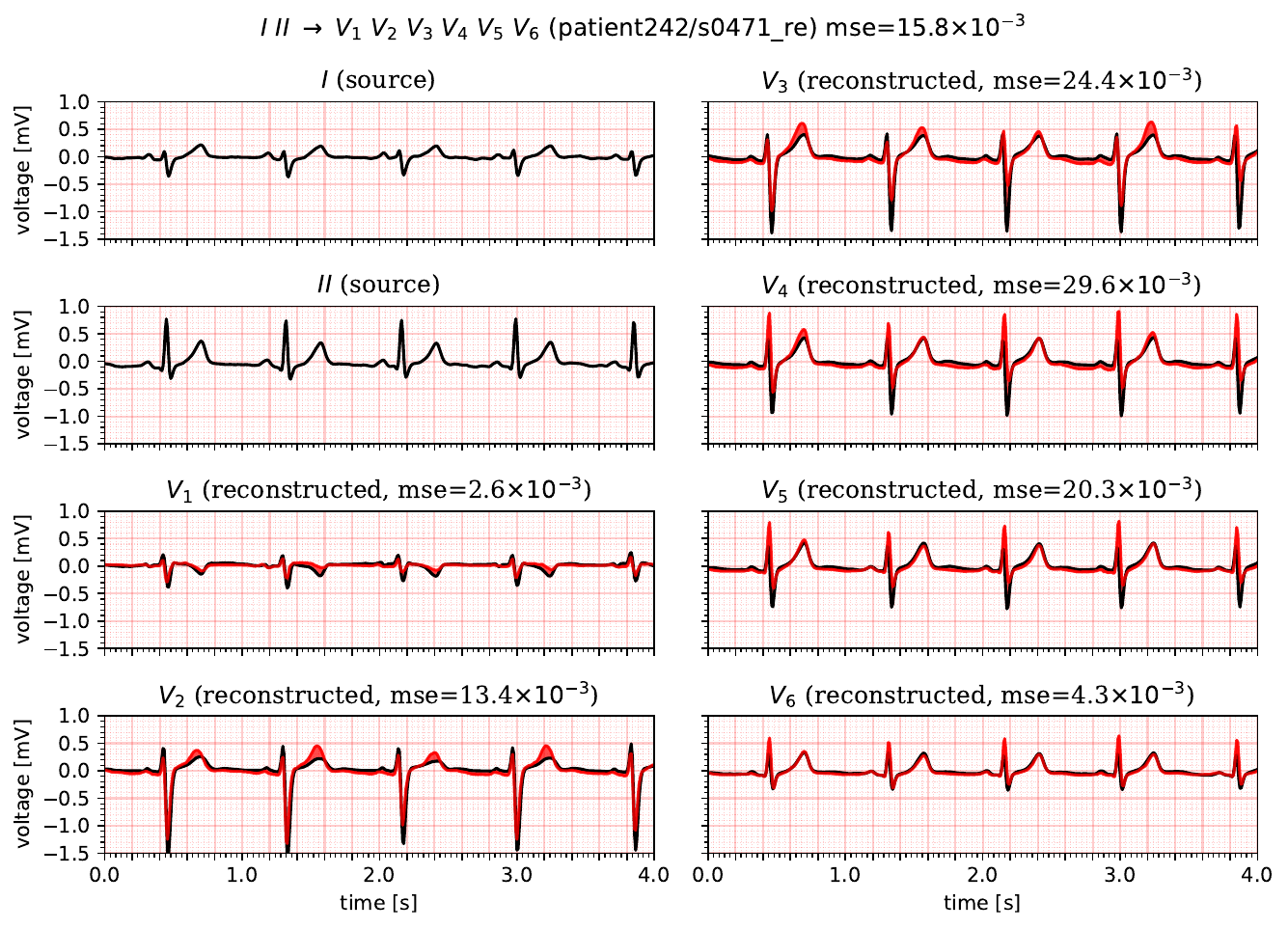}
    \caption{Patient \#242, female, 28 years old, healthy control. Reconstruction of precordial leads from the set of limb leads (I, II), exemplary ECG recording from the PTB dataset. Black solid lines represent the real signal; red lines represent the reconstructed signal, and the difference between them is filled with a light red. The MSE value for each reconstructed lead is shown at the top of the subplot. Only the first 4 seconds of the signal are shown.}\label{fig:base_ex1}
\end{figure}

\begin{figure}[h]
    \centering
    \includegraphics[width=\textwidth]{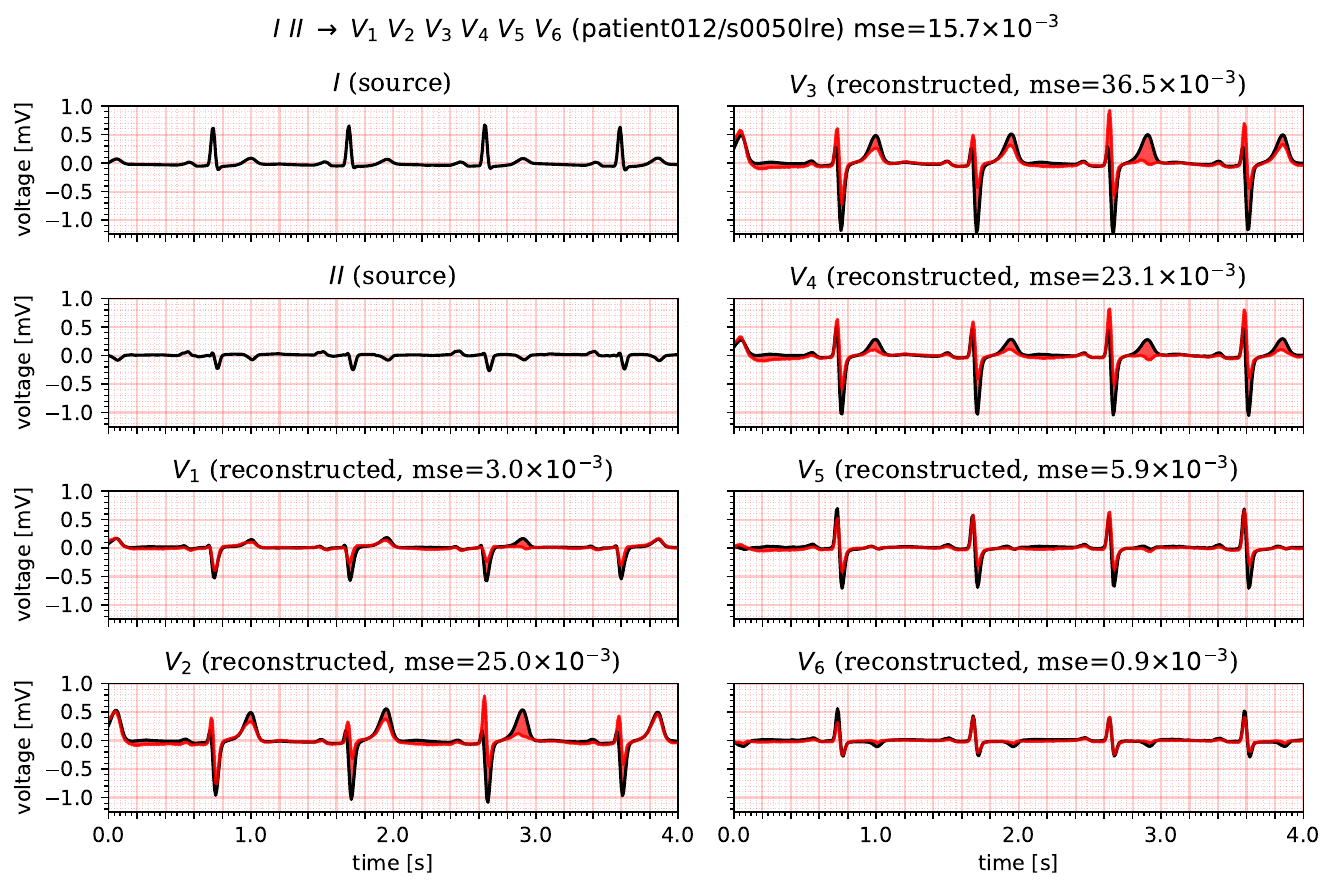}
    \caption{Patient \#012, male, 67 years old, reason for admission: myocardial infarction. See Fig. \ref{fig:base_ex1} caption for details.}\label{fig:base_ex2}
\end{figure}

\begin{figure}[h]
    \centering
    \includegraphics[width=\textwidth]{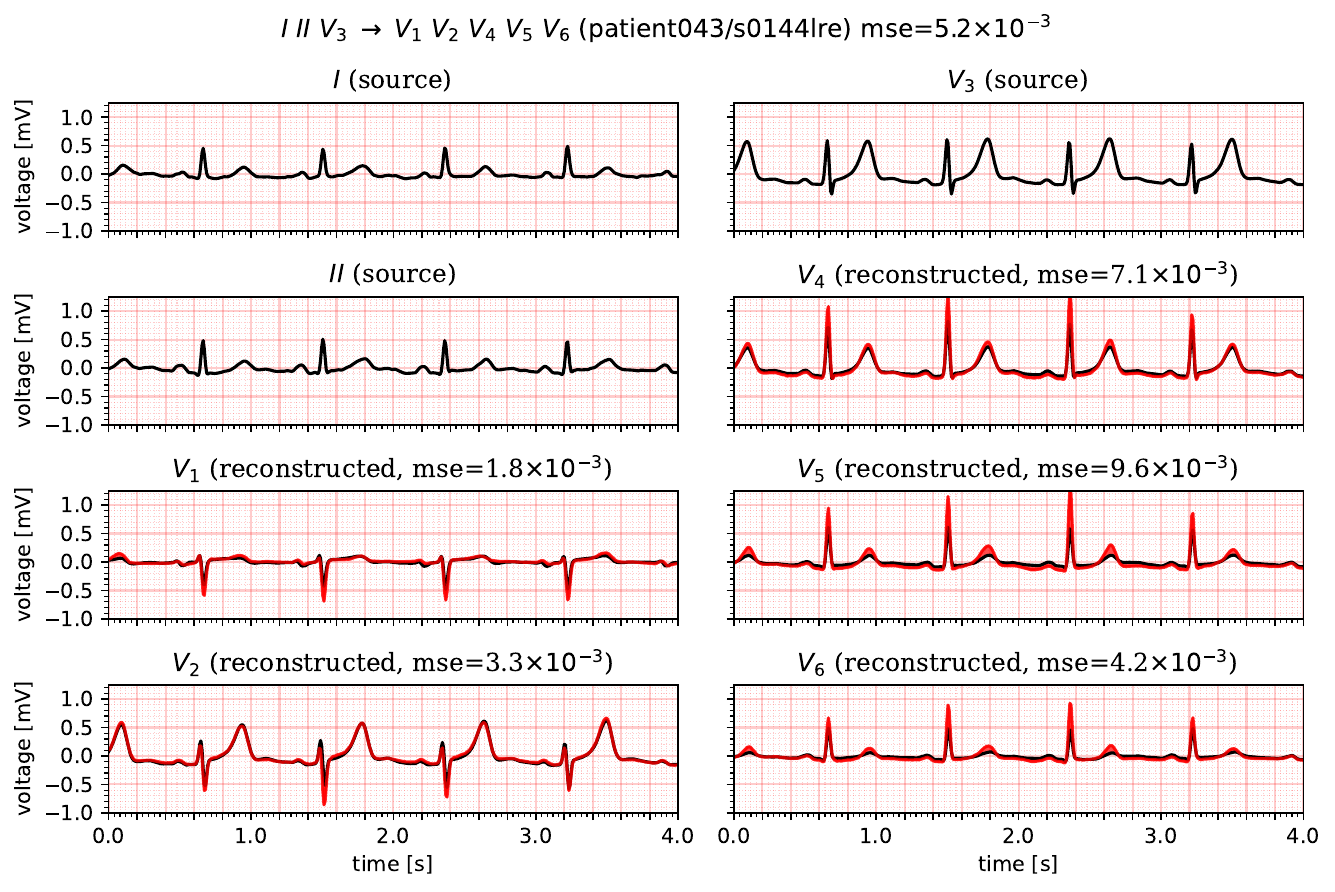}
    \caption{Patient \#43, female, 52 years old. Reason for admission: myocardial infarction. Reconstruction of precordial leads from the set of limb leads (I, II) and the V3 lead. See Fig. \ref{fig:base_ex1} caption for details.}\label{fig:v3_ex1}
\end{figure}

\begin{figure}[h]
    \centering
    \includegraphics[width=\textwidth]{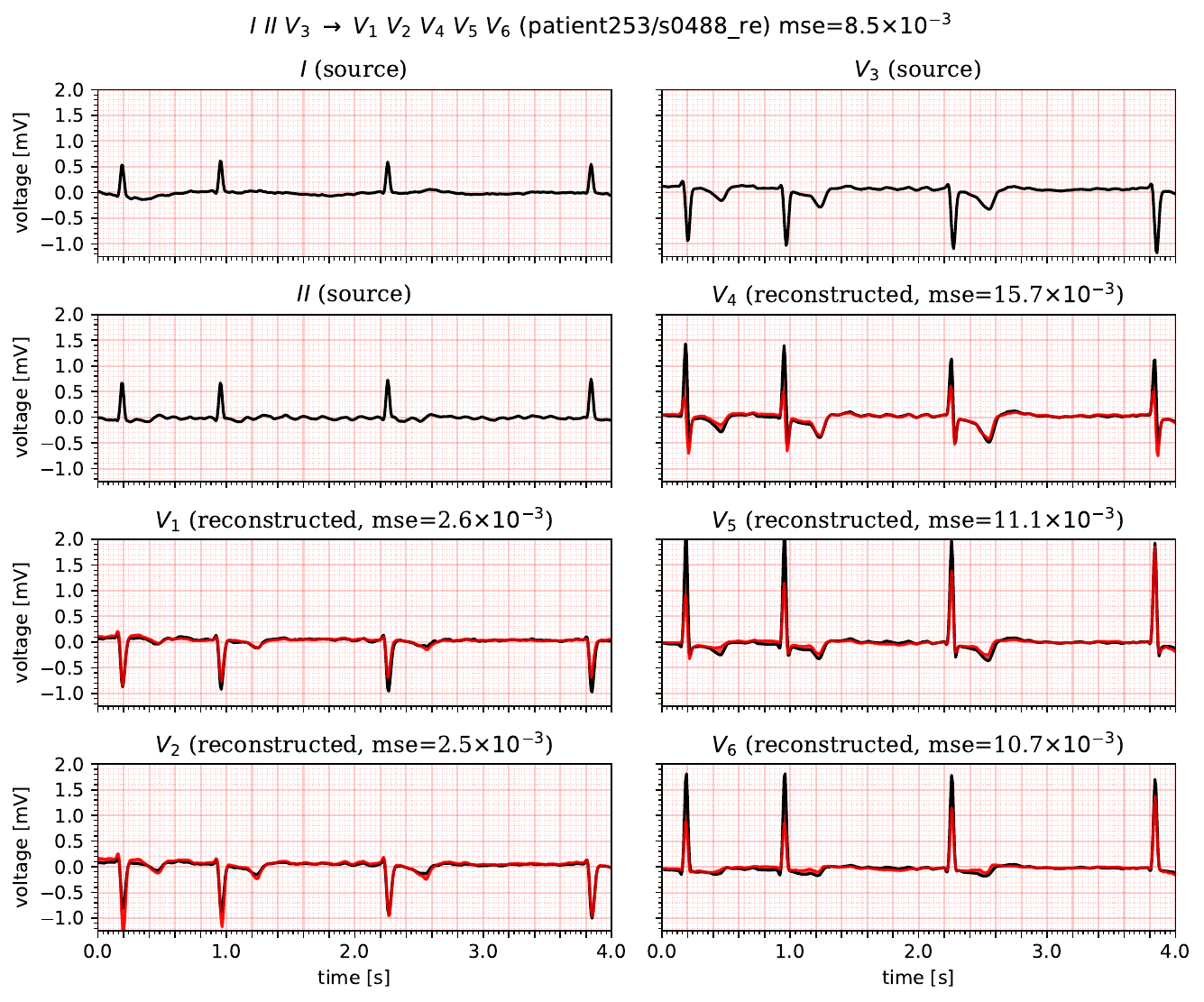}
    \caption{Patient \#253, male, 62 years old. Reason for admission: Cardiomyopathy. Bundle branch block diagnosed. Reconstruction of precordial leads from the set of limb leads (I, II) and the V3 lead. See Fig. \ref{fig:base_ex1} caption for details.}\label{fig:v3_ex2}
\end{figure}

A set of plots of input and reconstructed leads for the I+II and I+II+V3 models is shown in Fig. \ref{fig:base_ex1}-\ref{fig:v3_ex2}, to show the typical performance of the model. The primary source of error in the reconstruction is the T wave, which is the most volatile part of the ECG signal. The model struggles to accurately reconstruct the amplitudes of R waves, likely due to the low temporal resolution of the input data (125 Hz), which limits the ability to capture sharp signal transitions, and the fact that the standard MSE loss function underweights narrow, high-amplitude features like R waves. Future improvements could include adopting a weighted or morphology-aware loss function emphasizing diagnostically critical waveform components, such as QRS complexes.

\begin{figure}[h]
    \centering
    \includegraphics[width=\textwidth]{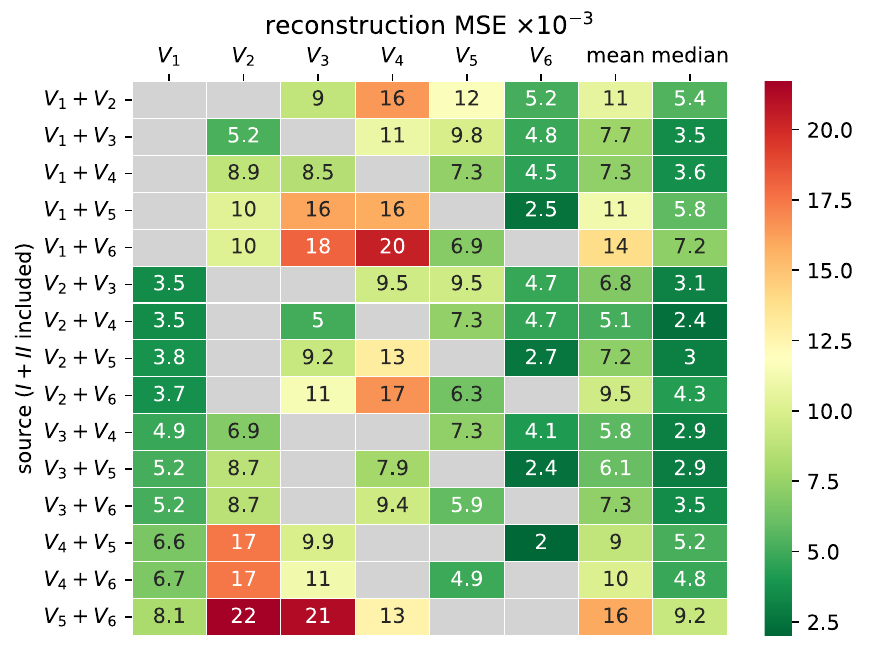}
    \caption{Performance of the models trained with two limb leads (I, II) and two precordial leads as input. The table shows the mean MSE between the reconstructed leads and the ground truth for each model. Mean and median MSE values for the testing dataset are shown at the most right columns.}\label{fig:double}
\end{figure}

\begin{figure}[h]
    \centering
    \includegraphics[width=\textwidth]{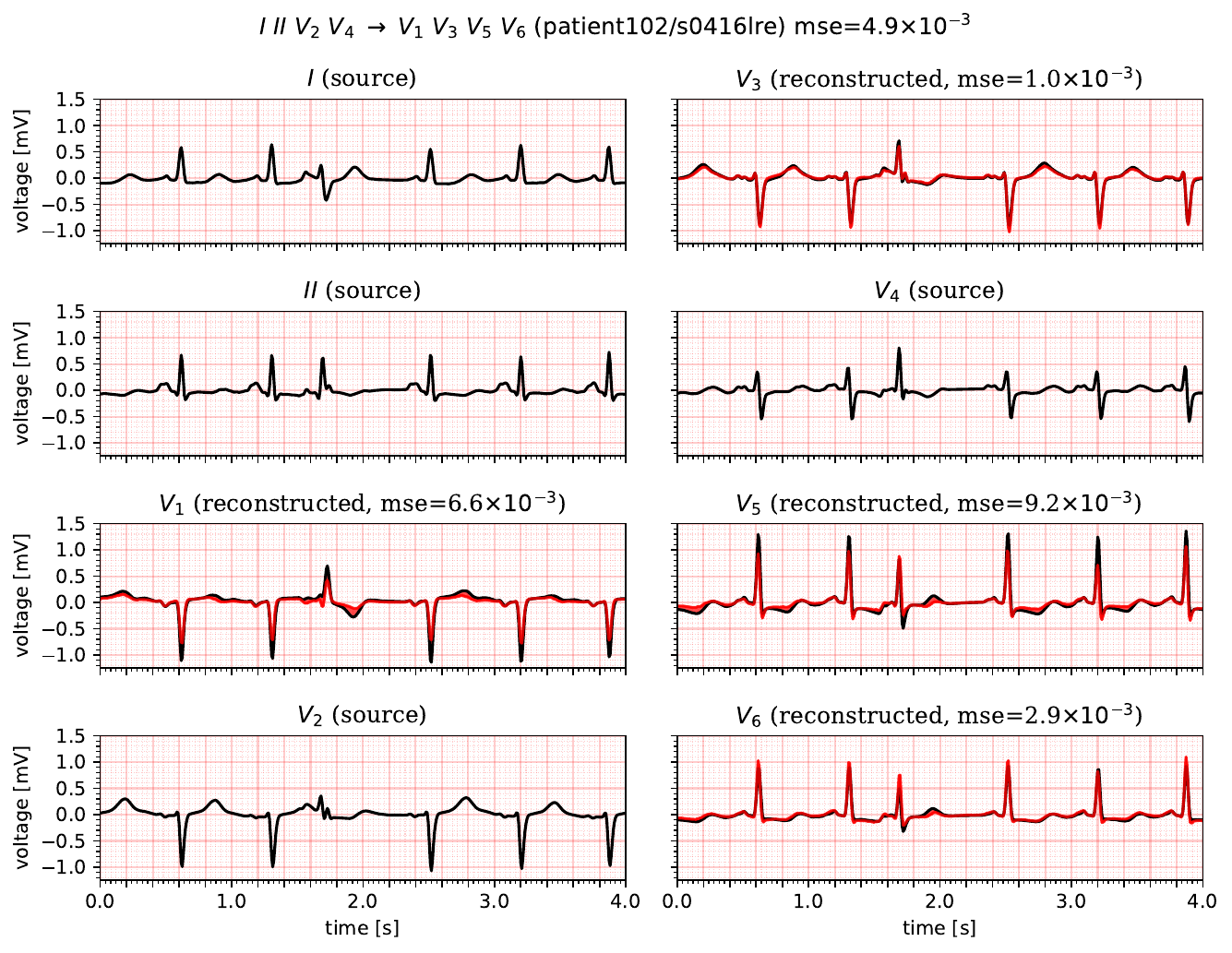}
    \caption{Patient \#102, female, 72 years old. Reason for admission: myocardial infarction. See Fig. \ref{fig:base_ex1} caption for details.}\label{fig:v2v4_ex1}
\end{figure}

Fig. \ref{fig:double} shows the results of the models trained with two limb leads and two precordial leads as input. Each combination of two precordial leads was examined. In this group, we observed significantly better reconstruction of the precordial leads than in the first group. The plots of the input and reconstructed leads for the selected model (I+II+V2+V4) are shown in Fig. \ref{fig:v2v4_ex1} for a selected example with a MSE value close to the average MSE of all signals for this model.

The PTB dataset used for testing predominantly consists of pathological cases: out of 268 patients, only 52 are from the healthy control group. The remaining cases include diagnoses such as myocardial infarction, cardiomyopathy, bundle branch block, dysrhythmias, myocardial hypertrophy, valvular heart disease, and myocarditis. Therefore, the average reconstruction results reported in this study primarily represent diseased subjects. Although we observed healthy ECG signals exhibited lower reconstruction errors than pathological ones, a detailed stratified analysis by specific disease classes was not performed. Such an analysis would be valuable in further characterizing model behavior across different cardiac conditions. Still, it is considered beyond the scope of the present work, particularly given the need to manage results across 25 different input configurations. Future studies could investigate class-specific reconstruction performance to optimize models for particular diagnostic categories.

\section{Discussion}

First, let us discuss the results of the single-lead models. In these models, we take only one limb lead as input and try to reconstruct all the precordial leads. The reconstruction quality can be interpreted as a measure of the correlation between the input and precordial leads. We must remember that all precordial leads have the same reference point, the Wilson central terminal, so the information from all the limb electrodes is present in all precordial leads to some extent. However, by comparing the single-lead models' results, we can also try to estimate the correlation between the single electrodes.

The results show that leads V1 and V6 are reconstructed with the highest accuracy in all three scenarios (Fig. \ref{fig:single}, first three rows). In particular, lead V6 appears more closely correlated with limb leads II and I. A slightly better reconstruction of V6 from lead II than from lead I suggests a stronger correlation between the potential on the V6 electrode and the potential on the LL electrode (left leg) than LA (left arm), since the electrode potential RA (right arm) is present in both leads I and II.

The V3 lead exhibited the lowest reconstruction accuracy in all three scenarios. This suggests that the potential on the V3 electrode carries information that is not present in the limb leads. This is consistent with that V3 is placed closest to the interventricular septum and is, on average, equally susceptible to potentials from both ventricles. Since it is centrally located on the chest, it captures a unique combination of signals not as prominent on the limb leads. The significance of the V3 lead will be seen in the 3-lead models.

To estimate the correlation between precordial leads, we trained models with two limb leads and one precordial lead as input (I + II + Vx). We also trained a model with two limb leads (I+II) as input, which serves as a baseline for the comparison.
The first observation is that precordial leads are the most correlated with the closest precordial leads, e.g., including the V2 lead in the input leads results in significantly better reconstruction of the V1 and V3 leads. This is very intuitive since the precordial electrodes are placed in a line along the heart’s axis, and the potential of one electrode is expected to be the most correlated with the potentials of the adjacent electrodes.
The results show that the reconstruction of the precordial leads is the best when the V3 lead is used as input. This confirms the previous observation that the V3 lead carries the most information about the heart’s electrical activity, which is not present in the limb leads. This observation aligns with the anatomical placement of V3 near the interventricular septum, which allows it to sample electrical activity from both ventricles. From a signal-processing perspective, its inclusion likely contributes orthogonal information that complements the limb leads. This reinforces the idea that lead configuration impacts the informational redundancy within ECG recordings, and has practical implications for lead selection in portable systems.

The worst reconstruction is obtained when the V6 lead is used as an input. This observation is consistent with the previous one, which states that the V6 lead is the most correlated with the limb leads and carries the least additional information. Including the V6 lead in the input leads does not affect the reconstruction of the V1-V4 leads and improves the reconstruction of only the V5 lead, measured from the closest electrode to the V6 electrode.

The following models assume a scenario where we want to reconstruct all the precordial leads from two limb leads and two precordial leads. This scenario is possible in a clinical setting, where precordial leads are unavailable due to the patient’s condition or technical limitations. A performant model in this scenario would open the way for developing a portable ECG device with fewer leads, which could be used in emergencies or remote areas. The decision on which ECG leads to include in such a device would be based on the correlation between the leads and the reconstruction quality. A naive approach would suggest that the best combination of leads should include lead V3, as it gave the best result in the previous models. However, the best combination of leads is  I + II + V2 + V4.
Leads V2 and V4 are positioned over areas central to the heart's electrical activity. V2 is placed in the fourth intercostal space at the left sternal border, while V4 is placed in the fifth intercostal space at the midclavicular line. The proximity of these leads to critical cardiac structures allows them to capture significant electrical activity from both the anterior and septal regions of the heart. V2 and V4 provide a balanced view of the heart’s electrical activity; V2 captures data from the right ventricle and the septal area, while V4 captures signals from the anterior wall of the left ventricle. This complementary information is crucial for accurately reconstructing the missing leads. Although adjacent leads (like V2 and V3) provide highly correlated data, they can introduce redundancy rather than additional helpful information. V2 and V4, being slightly more spatially separated, offer diverse and non-redundant information that enhances reconstruction accuracy.

From a systems integration perspective, the model is well-suited for real-time or point-of-care deployment. The U-net architecture, as implemented in this study, maintains moderate parameter count and computational complexity. The fixed-length input segments are small enough to allow rapid inference. These characteristics make the proposed model highly compatible with bedside monitors, portable ECG systems, and remote health monitoring devices. Future work may involve prototyping deployments on representative hardware platforms to benchmark inference speed and memory usage formally. One limitation of the current study is that the model was trained and evaluated exclusively on clinical-grade ECG recordings obtained from standard 12-lead systems (PTB-XL and PTB datasets). Wearable ECG devices, such as smartwatches or portable monitors, typically exhibit different noise profiles, including motion artifacts, lower signal-to-noise ratios, and electrode placement variability. As a result, direct generalization to wearable device signals is not guaranteed. To achieve optimal generalization, future work could incorporate domain adaptation strategies, such as fine-tuning on wearable device datasets or augmenting training data with synthetic noise, to explicitly model the noise characteristics of wearable signals. This would enhance the model's applicability to real-world telemedicine and mobile health scenarios.

\section{Conclusion}

The results of this study are promising and show the potential of deep learning models for the reconstruction of ECG. The method for estimating the amount of information carried by each lead and the correlations between the leads may be significant from the clinical point of view.
First, the proposed method can optimize ECG use. By quantifying the information carried by each lead, clinicians can prioritize specific leads over others. This can be especially useful when a full 12-lead ECG, such as in remote monitoring or emergencies, is impractical. Streamlining lead selection could improve diagnostic efficiency without compromising accuracy.
Second, the method can reduce redundancy. Correlations between leads can reveal which ones provide redundant information. By identifying this redundancy, clinicians can reduce the number of leads used in specific scenarios, potentially simplifying the equipment, reducing patient discomfort, and shortening ECG reading times.
Third, the method can improve personalized diagnostics. Different patients may have varying degrees of information spread across ECG leads. The presented method can help tailor the interpretation of the ECG to individual patients. It can also help detect subtle cardiac abnormalities.
Another value of the method is that it can provide a better understanding of cardiac physiology. Understanding the correlations between leads can offer more profound insight into the heart's electrical activity and how it propagates across different axes, potentially aiding in a more accurate localization of ischemic or arrhythmic areas.
Finally, the method can improve telemedicine. For remote health monitoring, using fewer leads without sacrificing the quality of information could be a game-changer. It could make ECG monitoring more accessible and affordable, especially in low-resource settings.
These advances can lead to faster diagnosis, more targeted treatment, and potentially lower healthcare costs \cite{Siontis2021}.

 \bibliographystyle{elsarticle-num} 
 \bibliography{main}

\begin{thebibliography}{10}
\expandafter\ifx\csname url\endcsname\relax
  \def\url#1{\texttt{#1}}\fi
\expandafter\ifx\csname urlprefix\endcsname\relax\def\urlprefix{URL }\fi
\expandafter\ifx\csname href\endcsname\relax
  \def\href#1#2{#2} \def\path#1{#1}\fi

\bibitem{Yoo2023}
H.~Yoo, Y.~Yum, Y.~Kim, J.-H. Kim, H.-J. Park, H.~J. Joo, Restoration of missing or low-quality 12-lead ecg signals using ensemble deep-learning model with optimal combination, Biomedical Signal Processing and Control 83 (2023) 104690.
\newblock \href {https://doi.org/10.1016/j.bspc.2023.104690} {\path{doi:10.1016/j.bspc.2023.104690}}.

\bibitem{Xue2024}
J.~Xue, Does a reduced ecg lead set contain the full 12-lead ecg information for interpretation, in: 2024 Computing in Cardiology Conference (CinC), Vol.~51 of CinC2024, Computing in Cardiology, 2024.
\newblock \href {https://doi.org/10.22489/cinc.2024.473} {\path{doi:10.22489/cinc.2024.473}}.

\bibitem{Pipberger1961-fi}
H.~V. Pipberger, S.~M. Bialek, J.~K. Perloff, H.~W. Schnaper, Correlation of clinical information in the standard 12-lead {ECG} and in a corrected orthogonal 3-lead {ECG}, Am. Heart J. 61~(1) (1961) 34--43.
\newblock \href {https://doi.org/10.1016/0002-8703(61)90514-2} {\path{doi:10.1016/0002-8703(61)90514-2}}.

\bibitem{Holderith2018-ft}
M.~Holderith, T.~Schanze, {Cross-Correlation} based comparison between the conventional 12-lead {ECG} and an {EASI} derived 12-lead {ECG}, Curr. Dir. Biomed. Eng. 4~(1) (2018) 621--624.
\newblock \href {https://doi.org/10.1515/cdbme-2018-0149} {\path{doi:10.1515/cdbme-2018-0149}}.

\bibitem{Jain2023-rg}
U.~Jain, A.~Butchy, M.~Leasure, V.~Covalesky, D.~McCormick, G.~Mintz, Redundancy and novelty between {ECG} leads based on linear correlation, in: Proceedings of the 16th International Joint Conference on Biomedical Engineering Systems and Technologies, SCITEPRESS - Science and Technology Publications, 2023, pp. 359--365.
\newblock \href {https://doi.org/10.5220/0011815700003414} {\path{doi:10.5220/0011815700003414}}.

\bibitem{Zhang2022-le}
C.~Zhang, J.~Li, S.~Pang, F.~Xu, S.~Zhou, A 12-lead {ECG} correlation network model exploring the inter-lead relationships, EPL 140~(3) (2022) 31001.
\newblock \href {https://doi.org/10.1209/0295-5075/ac9b89} {\path{doi:10.1209/0295-5075/ac9b89}}.

\bibitem{Lence2025}
A.~Lence, F.~Granese, A.~Fall, B.~Hanczar, J.-E. Salem, J.-D. Zucker, E.~Prifti, Ecgrecover: A deep learning approach for electrocardiogram signal completion, in: Proceedings of the 31st ACM SIGKDD Conference on Knowledge Discovery and Data Mining V.1, KDD ’25, ACM, 2025, p. 2359–2370.
\newblock \href {https://doi.org/10.1145/3690624.3709405} {\path{doi:10.1145/3690624.3709405}}.

\bibitem{Frank1956}
E.~Frank, An accurate, clinically practical system for spatial vectorcardiography, Circulation 13~(5) (1956) 737–749.
\newblock \href {https://doi.org/10.1161/01.cir.13.5.737} {\path{doi:10.1161/01.cir.13.5.737}}.

\bibitem{Gargiulo}
G.~D. Gargiulo, True unipolar {ECG} machine for wilson central terminal measurements, Biomed Res. Int. 2015 (2015) 586397.
\newblock \href {https://doi.org/10.1155/2015/586397} {\path{doi:10.1155/2015/586397}}.

\bibitem{Gargiulo2016}
G.~Gargiulo, P.~Bifulco, M.~Cesarelli, A.~McEwan, A.~OLoughlin, J.~Tapson, A.~Thiagalingam, The wilson's central terminal ({WCT)}: A systematic error in {ECG} recordings, Heart Lung Circ. 25 (2016) S263.
\newblock \href {https://doi.org/10.1016/j.hlc.2016.06.613} {\path{doi:10.1016/j.hlc.2016.06.613}}.

\bibitem{Moeinzadeh}
H.~Moeinzadeh, P.~Bifulco, M.~Cesarelli, A.~L. McEwan, A.~O'Loughlin, I.~M. Shugman, J.~C. Tapson, A.~Thiagalingam, G.~D. Gargiulo, Minimization of the wilson's central terminal voltage potential via a genetic algorithm, BMC Res. Notes 11~(1) (Dec. 2018).
\newblock \href {https://doi.org/10.1186/s13104-018-4017-y} {\path{doi:10.1186/s13104-018-4017-y}}.

\bibitem{Man2015}
S.~Man, A.~C. Maan, M.~J. Schalij, C.~A. Swenne, Vectorcardiographic diagnostic \& prognostic information derived from the 12-lead electrocardiogram: Historical review and clinical perspective, J. Electrocardiol. 48~(4) (2015) 463--475.
\newblock \href {https://doi.org/10.1016/j.jelectrocard.2015.05.002} {\path{doi:10.1016/j.jelectrocard.2015.05.002}}.

\bibitem{Boonstra}
M.~J. Boonstra, D.~H. Brooks, P.~Loh, P.~M. van Dam, {CineECG}: A novel method to image the average activation sequence in the heart from the 12-lead {ECG}, Comput. Biol. Med. 141~(105128) (2022) 105128.
\newblock \href {https://doi.org/10.1016/j.compbiomed.2021.105128} {\path{doi:10.1016/j.compbiomed.2021.105128}}.

\bibitem{Schmitt}
O.~H. Schmitt, R.~B. Levine, E.~Simonson, Electrocardiographic mirror pattern studies. {I}, Am. Heart J. 45~(3) (1953) 416--428.
\newblock \href {https://doi.org/10.1016/0002-8703(53)90152-5} {\path{doi:10.1016/0002-8703(53)90152-5}}.

\bibitem{Brody}
D.~A. Brody, G.~D. Copeland, Electrocardiographic cancellation: Some observations concerning the ``nondipolar'' fraction of precordial electrocardiograms, Am. Heart J. 56~(3) (1958) 381--395.
\newblock \href {https://doi.org/10.1016/0002-8703(58)90277-1} {\path{doi:10.1016/0002-8703(58)90277-1}}.

\bibitem{Vaidya}
G.~N. Vaidya, S.~Antoine, S.~H. Imam, H.~Kozman, H.~Smulyan, D.~Villarreal, Reciprocal {ST-segment} changes in myocardial infarction: Ischemia at distance versus mirror reflection of {ST-elevation}, Am. J. Med. Sci. 355~(2) (2018) 162--167.
\newblock \href {https://doi.org/10.1016/j.amjms.2017.09.004} {\path{doi:10.1016/j.amjms.2017.09.004}}.

\bibitem{Wang}
J.~Wang, J.~Li, S.~Diao, H.~Xu, F.~Ding, Atypical de winter {ECG} pattern may be the mirror image of {ST} elevation, Ann. Noninvasive Electrocardiol. 27~(3) (2022) e12915.
\newblock \href {https://doi.org/10.1111/anec.12915} {\path{doi:10.1111/anec.12915}}.

\bibitem{Meek}
S.~Meek, {ABC} of clinical electrocardiography: Introduction. {II---Basic} terminology, BMJ 324~(7335) (2002) 470--473.
\newblock \href {https://doi.org/10.1136/bmj.324.7335.470} {\path{doi:10.1136/bmj.324.7335.470}}.

\bibitem{Riera}
A.~R. P{\'e}rez-Riera, R.~Barbosa-Barros, R.~Daminello-Raimundo, L.~C. de~Abreu, Main artifacts in electrocardiography, Ann. Noninvasive Electrocardiol. 23~(2) (2018) e12494.
\newblock \href {https://doi.org/10.1111/anec.12494} {\path{doi:10.1111/anec.12494}}.

\bibitem{Buchner}
T.~Buchner, M.~Zajdel, K.~P\k{e}czalski, P.~Nowak, Finite velocity of {ECG} signal propagation: preliminary theory, results of a pilot experiment and consequences for medical diagnosis, Sci. Rep. 13~(1) (2023) 4716.
\newblock \href {https://doi.org/10.1038/s41598-023-29904-2} {\path{doi:10.1038/s41598-023-29904-2}}.

\bibitem{unet}
O.~Ronneberger, P.~Fischer, T.~Brox, U-net: Convolutional networks for biomedical image segmentation, in: N.~Navab, J.~Hornegger, W.~M. Wells, A.~F. Frangi (Eds.), Medical Image Computing and Computer-Assisted Intervention -- MICCAI 2015, Springer International Publishing, Cham, 2015, pp. 234--241.
\newblock \href {https://doi.org/10.1007/978-3-319-24574-4\_28} {\path{doi:10.1007/978-3-319-24574-4\_28}}.

\bibitem{autoencoders}
I.~Goodfellow, Y.~Bengio, A.~Courville, Deep learning. Adaptive computation and machine learning, Mass: The MIT press, Cambridge, 2016, Ch.~14, pp. 499--523.

\bibitem{ptbxl1}
P.~Wagner, N.~Strodthoff, R.-D. Bousseljot, W.~Samek, T.~Schaeffter, \href{https://physionet.org/content/ptb-xl/1.0.3/}{Ptb-xl, a large publicly available electrocardiography dataset} (2022).
\newblock \href {https://doi.org/10.13026/KFZX-AW45} {\path{doi:10.13026/KFZX-AW45}}.
\newline\urlprefix\url{https://physionet.org/content/ptb-xl/1.0.3/}

\bibitem{ptbxl2}
P.~Wagner, N.~Strodthoff, R.-D. Bousseljot, D.~Kreiseler, F.~I. Lunze, W.~Samek, T.~Schaeffter, {PTB-XL}, a large publicly available electrocardiography dataset, Sci. Data 7~(1) (2020) 154.

\bibitem{ptb}
R.~Bousseljot, D.~Kreiseler, A.~Schnabel, Nutzung der ekg-signaldatenbank cardiodat der ptb \"{u}ber das internet (Jul. 2009).
\newblock \href {https://doi.org/10.1515/bmte.1995.40.s1.317} {\path{doi:10.1515/bmte.1995.40.s1.317}}.

\bibitem{physionet}
A.~L. Goldberger, L.~A. Amaral, L.~Glass, J.~M. Hausdorff, P.~C. Ivanov, R.~G. Mark, J.~E. Mietus, G.~B. Moody, C.~K. Peng, H.~E. Stanley, {PhysioBank}, {PhysioToolkit}, and {PhysioNet}: components of a new research resource for complex physiologic signals, Circulation 101~(23) (2000) E215--20.

\bibitem{Siontis2021}
K.~C. Siontis, P.~A. Noseworthy, Z.~I. Attia, P.~A. Friedman, Artificial intelligence-enhanced electrocardiography in cardiovascular disease management, Nature Reviews Cardiology 18~(7) (2021) 465–478.
\newblock \href {https://doi.org/10.1038/s41569-020-00503-2} {\path{doi:10.1038/s41569-020-00503-2}}.

\end{thebibliography}

\end{document}